\documentclass[prl,twocolumn,groupedaddress,preprintnumbers,showpacs]{revtex4}
\usepackage[T1]{fontenc}
\usepackage{graphicx}% Include figure files
\usepackage{dcolumn}% Align table columns on decimal point

\usepackage{color,graphics,graphicx,fancybox,fancyhdr}
\usepackage{latexsym}
\usepackage{amsmath}
\usepackage{amssymb}

\begin{document}
\bibliographystyle{apsrev}

\graphicspath{../Latex} \DeclareGraphicsExtensions{.eps,.ps}

\title{Experimental evidence of $s$-wave superconductivity in bulk CaC$_{6}$}

\author{G. Lamura$^{a}$, M. Aurino, G. Cifariello, E. Di Gennaro, and A. Andreone}

\affiliation{CNR-INFM Coherentia and Department of Physics,
University of Naples "Federico II", I-80125, Napoli, Italy.}

\author{N. Emery, C. Hérold, J.-F. Marêché, and P. Lagrange}

\affiliation{Laboratoire de Chimie du Solide Minéral-UMR 7555,
Université Henri Poincaré Nancy I, B.P. 239, 54506 Vand\oe
uvre-lès-Nancy Cedex, France.}

\date{\today}
\begin{abstract}

The temperature dependence of the in-plane magnetic penetration
depth, $\lambda_{ab}(T)$, has been measured in a $c$-axis oriented
polycrystalline CaC$_{6}$ bulk sample using a high-resolution
mutual inductance technique. A clear exponential behavior of
$\lambda_{ab}(T)$ has been observed at low temperatures, strongly
suggesting isotropic $s$-wave pairing. Data fit using the standard
BCS theory yields $\lambda_{ab}(0)=(720\pm 80)$ Å and
$\Delta(0)=(1.79\pm 0.08)$ meV. The ratio
$2\Delta(0)/k_{_B}T_{c}=(3.6\pm 0.2)$ gives indication for a
conventional weakly coupled superconductor.

\end{abstract}

%insert suggested PACS numbers in braces on next line
\pacs{74.25.Nf, 74.70.Ad, 81.05.Uw}

\maketitle

The recent discovery of superconductivity at 6.5 K and 11.5 K in
the graphite intercalation compounds (GICs) YbC$_{6}$
\cite{WELLER} and CaC$_{6}$ \cite{EMERY, WELLER} respectively has
renewed the theoretical interest in the physical properties of
this class of materials \cite{CSANYI,MAZIN,CALANDRA}. Graphite
intercalated with alkali-metals was known to undergo a
normal-to-superconducting transition since 1965 \cite{HANNAY},
with T$_{c}$ increasing with a larger alkali-metal concentration
(highest T$_{c}$ of 5 K for NaC$_{2}$ \cite{BELASH}).
Superconductivity in GICs with rare-earth metals was first
discovered with the synthesis of YbC$_{6}$. CaC$_{6}$ firstly
showed traces of superconductivity in a reduced quality sample
\cite{WELLER}, and only recently a clear transition has been put
in evidence in high quality bulk samples \cite{EMERY}. These
results open new perspectives in the physics of graphite. Its low
conductivity can be greatly enhanced by a large number of
reagents, continuously changing from a semimetallic to a good
metallic behavior. The discovery of superconductivity at easily
reachable temperatures can be helpful in understanding the
correlation between the charge transfer to the graphene layers and
T$_{c}$. It has been also proposed \cite{KLAPWIJK} that a deeper
comprehension of the pairing mechanism in GICs can shed a light on
the intrinsic or proximity-induced superconductivity reported in
carbon nanotubes \cite{KASUMOV}, which are at the core of
nanotechnology research.

The atomic structure of first stage GICs with metals usually
consists of a stacking of graphene sheets (ab plane) arranged in a
hexagonal configuration, with the intercalated atom occupying
interlayer sites above the centres of the hexagons. As shown by
Emery et al. \cite{EMERYII}, CaC$_{6}$ is the only member of the
MC$_{6}$ metal-graphite compounds exhibiting a rhombohedral
symmetry.

One of the main open questions making GICs an interesting class of
superconducting materials is related to the nature of the pairing
mechanism, whether it is driven by an ordinary electron-phonon
interaction \cite{MAZIN,CALANDRA} or due to electronic
correlations \cite{CSANYI}. The possibility of an unconventional,
excitonic or plasmonic, origin of superconductivity in GICs has
been invoked because of the nature of the energy bands in these
compounds. The intercalant atoms act as donors, thus producing a
charge transfer to the carbon layers. This results in partially
filled, mostly 2D in character, graphite $\pi$ bands. In addition,
in all compounds exhibiting superconductivity an interlayer 3D
$s$-band, well separated from the graphene sheets and formed by
nearly-free electrons propagating in the interstitial space,
crosses the Fermi surface and hybridizes with the $\pi$ bands. It
is interesting to note that the stronger is the hybridisation the
higher is T$_{c}$ \cite{CSANYI}. A sandwich structure consisting
of alternate layers of metal and semiconductor has been suggested
as a favorable environment for the excitonic mechanism since the
metal "free" electrons can tunnel into the gap region of the
semiconductor and interact with the excitons \cite{ALLENDER}.
Low-energy plasmons have been proposed as the dominant
contribution to superconductivity in metal-intercalated halide
nitrides \cite{BILL}. The recent analysis by Calandra and Mauri
\cite{CALANDRA} and Mazin and Molodotsov \cite{MAZIN2}, however,
points out that a simple electron-phonon interaction between the
intercalant $s$-band electrons and Ca in-plane and C out-of-plane
phonons in CaC$_{6}$ and Yb phonons in YbC$_{6}$ may be sufficient
to explain superconductivity in these GICs.

All that said, it seems clear that the experimental challenge for
understanding the origin of superconductivity in these compounds
should focus on the mechanism determining the pairing and on the
role played by the interlayer $s$-band. A first step in answering
these questions is to determine the symmetry of the
superconducting gap function and the nature of the elementary
excitations. The magnetic penetration depth $\lambda$ is known to
be a very sensitive probe of the low-lying quasiparticles energy,
and it is capable to give information which are significant on the
$\lambda$(0) scale rather than on the coherence length $\xi$(0)
scale, as it occurs in other spectroscopic tools. This corresponds
to probe the true "bulk" properties of a homogeneous
superconductor. To this aim, we have performed the first
high-resolution measurement of the in-plane magnetic penetration
depth $\lambda_{ab}(T)$ on a $c$-axis grown polycrystalline sample
of CaC$_{6}$. We find clear evidence of an exponentially activated
behavior of $\lambda_{ab}(T)$, consistent with an $s$-wave
symmetry of the gap function. In particular, the gap deduced from
the data fit is in full agreement with the BCS weak coupling value
(3.52), thus supporting a phonon-mediated nature of
superconductivity in this compound.

\begin{figure}[h]
\includegraphics[scale=0.5]{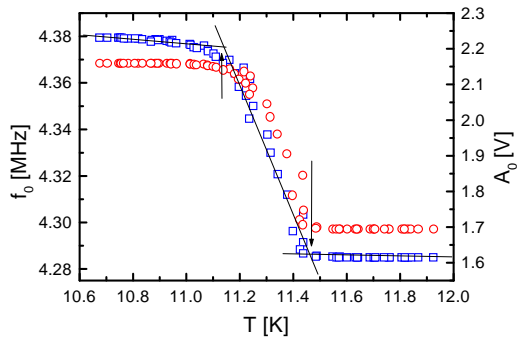}
\caption{$T_{c}$ inductive characterization of CaC$_{6}$. At the
superconducting transition a large change in the resonant
frequency $f_{0}$ (\small{$\square$}\normalsize) and amplitude
$A_{0}$ (\Large{$\circ$}\normalsize) of the oscillating signal is
observed. The arrows indicate the transition width $\Delta T_{c}$.
The solid lines are a guide for the eye.} \label{Figure1}
\end{figure}

Bulk CaC$_{6}$ has been synthesized from highly oriented pyrolytic
graphite \cite{EMERY, EMERYII}. The reaction is carried out for
ten days between a pyrolytic graphite platelet and a molten
lithium-calcium alloy at around 350$^{\circ}$C, under very pure
argon atmosphere. The reactive alloy has to be very rich in
lithium, with a composition between 70 and 80 at.\% of Li. Despite
such a low calcium concentration, no lithium is present in the
final reaction product and calcium alone is intercalated into
graphite. For a more detailed description of the technique see
\cite{EMERYII} and references therein. The resulting samples are
platelike polycristals with the $c$-axes of all the crystallites
forming the highly oriented graphite parallel to each other,
whereas in the perpendicular plane the material is disordered,
leading to an average of $a$ and $b$ directions, denoted as $ab$.
The data presented here have been taken on a sample having a
roughly rectangular shape of about 2.5x2.5 mm$^{2}$ and thickness
of 0.1 mm. The as grown samples are shiny and silver in color but
tarnish quickly in air. To ensure that the analysis is performed
on a clean and non-reacted surface, we have studied the same
sample before and immediately after cleaving it. We have measured
the in-plane magnetic penetration depth $\lambda_{ab}(T)$ in the
range 1.8 K - $T_{c}$ by using a single-coil mutual inductance
technique described in detail elsewhere \cite{GAUZZI}. Typical
frequency and magnitude of the inducing field B$_{ac}$
perpendicular to the film surface are 2-4 MHz and $<$ 0.1 mT
respectively. Care has been taken to always operate in the linear
response regime, monitored by varying the applied magnetic field.
No significant edge effects are expected in samples equal or
larger than two times the coil diameter $d$, as in the present
case ($d$=0.8 mm) \cite{LAMURAIII}.

\begin{figure}[h]
\includegraphics[scale=0.51]{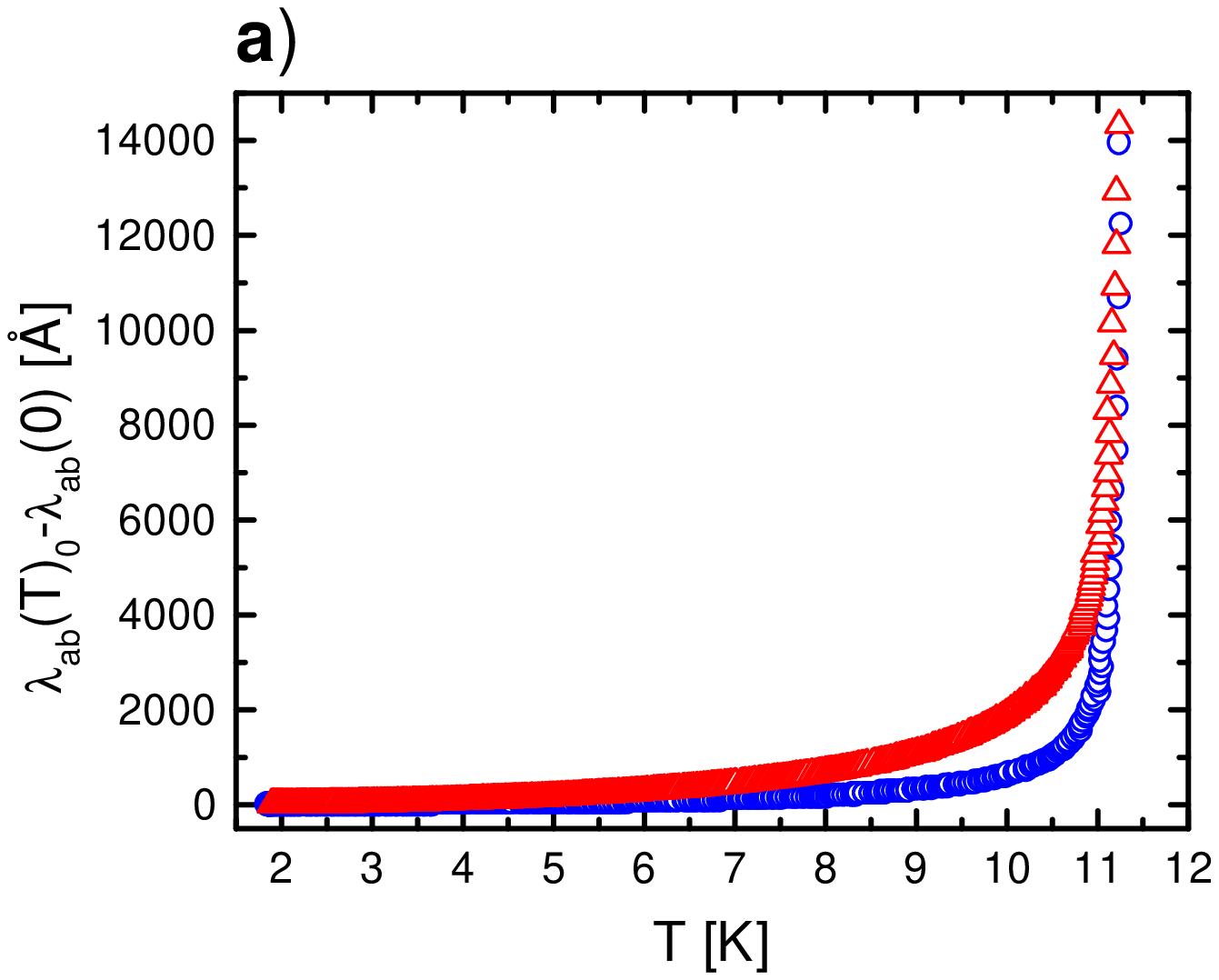}
\includegraphics[scale=0.68]{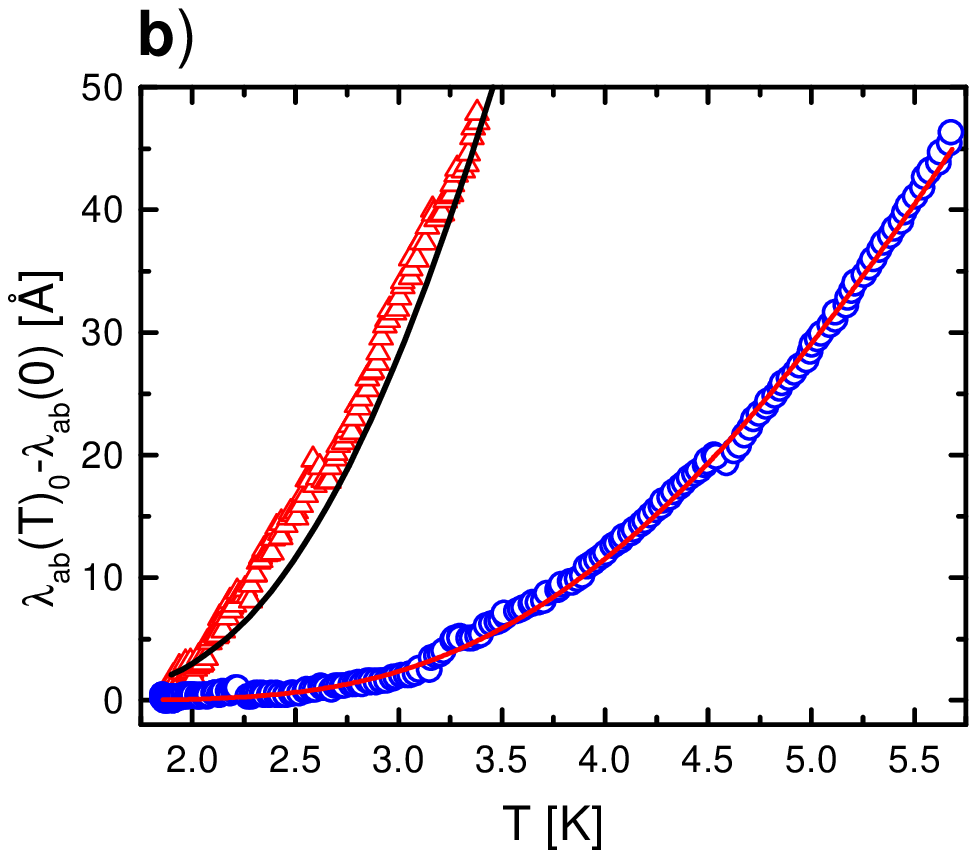}
\caption{a) Variation of the in-plane magnetic penetration depth
up to T$_{c}$ for the sample under test before
(\small{$\bigtriangleup$}\normalsize) and after cleavage
(\Large{$\circ$}\normalsize). b) The same as in fig. 2a, but at
low temperatures ($T<T_{c}/2$). The solid lines represent the BCS
fits (see text).} \label{Figure2}
\end{figure}

In Fig. 1, we report the inductive characterization of the
critical temperature by observing the behavior of the resonant
frequency $f_{0}$ and the signal amplitude $A_{0}$ in the
transition region \cite{GAUZZI}. The onset of superconductivity is
at $T_{c}^{on}=(11.46\pm0.05)$ K with a transition width $\Delta
T_{c}=(0.40\pm0.05)$ K. No difference has been found either in
$T_{c}^{on}$ or in $\Delta T_{c}$ before and after cleaving the
sample. In fig. 2a we show the change of the in-plane magnetic
penetration depth in the overall temperature range for the sample
under test before and after cleavage. A much larger variation is
observed in the former case, in comparison with the freshly
cleaved sample. At low temperatures (fig. 2b) this feature is even
more evident.

\begin{figure}[h]
\includegraphics[scale=0.5]{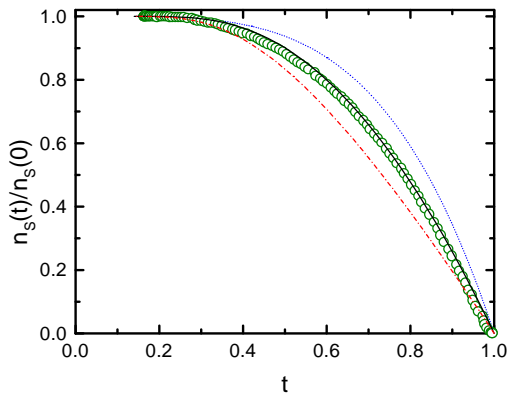}
\caption{Normalized superfluid density versus the reduced
temperature for the cleaved sample (\Large{$\circ$}\normalsize).
The short dash-dotted and the continuous line represent the weakly
coupled BCS model in the clean and in the dirty limit
respectively. The dotted line shows the two fluids
behavior.}\label{Figure3}
\end{figure}

We firstly analyze the above results at low temperatures within
the framework of a standard BCS $s$-wave model. According to the
theory, the temperature dependence of $\Delta\lambda_{ab}$ is
proportional to the fraction of normal electrons in the low
temperature limit (T<$T_{c}$/2) and follows a thermally activated
behavior given by \cite{TURNEAURE}:

\begin{eqnarray}\label{1}
\lambda(T)-\lambda(0)=
\lambda(0)\sqrt{\frac{\pi\Delta(0)}{2k_{_B}T}}\cdot
\exp\left({-\frac{\Delta(0)}{k_{_B}T}}\right)
\end{eqnarray}
where $\Delta(0)$ is the zero-temperature superconducting gap. In
fig. 2b we report also the result of a fit procedure performed by
using eq. (1) on the measurements carried out before and after
cleavage (solid lines). In both cases the BCS single-gap model
well describes the data. In particular, for the freshly cleaved
sample the fit is extremely good, yielding the following
superconducting parameters: $\Delta(0)=(1.79\pm0.08)$ meV and
$\lambda_{ab}(0)=(720\pm80)$ Å. The ratio $2\Delta(0)/k_{_B}T_{c}$
is then evaluated to be $(3.6\pm0.2)$. Before the cleavage, the
same fit on the sample data gives a zero-temperature penetration
depth almost doubled and a zero-temperature gap lowered by 30\%.
Such a difference in the behavior is very likely due to the
presence of a thin (on the scale of $\lambda$(0)) reacted layer on
the uncleaved sample surface, having depressed superconducting
properties. This is confirmed by the fact that the inductively
measured superconducting "bulk" transition is not affected by the
degraded layer. For this reason, the discussion on the
experimental data will focus only on the freshly cleaved sample.
Using the results of the BCS fit at low temperatures, we have then
tried to deduce the London penetration depth in the ab plane over
the whole temperature range from the relation \cite{TINK}:

\begin{eqnarray}\label{1}
\lambda_{L}(T)=\lambda_{eff}(T) \left[ 1+
\frac{\xi_{ab}(0)}{J(0,T)\cdot l_{m.f.p.} } \right]^{-1/2}
\end{eqnarray}

where $\xi_{ab}(0)$ is the zero temperature in-plane coherence
length, $l_{m.f.p.}(0)$ is the mean free path, and $J(0,T)$ is the
real-space kernel valid for a local electrodynamic response. The
local limit can be safely used because $\xi_{ab}(0)=350$ Å <
$\lambda(0)$, by magnetization measurements performed in samples
from the same batch \cite{EMERY}. The fit shows that the screening
response of the CaC$_{6}$ sample is definitely in the dirty limit
($l_{m.f.p.}(0)$ < $\xi_{ab}(0)$). This can be also seen plotting
the normalized superfluid density $n_{s}(T)/n_{s}(0) =
[\lambda(0)/\lambda(T)]^{2}$ (fig. 3). The experimental curve lies
just between the clean local BCS limit and the two fluid behavior
and it is well described by the BCS calculation in the dirty limit
\cite{TINK} using the following equation:

\begin{eqnarray}\label{1}
\left[\frac{\lambda(0)}{\lambda(T)}\right]^{2}=\frac{\Delta(T)}{\Delta(0)}\tanh\left(\frac{\Delta(T)}{2k_{_{B}}T}
\right)
\end{eqnarray}

This procedure allows us to confirm that the electrodynamic
response of CaC$_{6}$ follows a single-gap s-wave behavior
throughout the entire temperature range. However, nothing can be
reliably said on the value of the mean free path and of the
zero-temperature London penetration depth $\lambda_{L}(0)$. A
precise estimation of $l_{m.f.p.}(0)$ confirming our conclusions
must await an independent experiment. All that we can say is that,
very likely, the dirty limit response is due to the presence of a
certain amount of disorder in the Ca distribution during the
intercalation process \cite{DRESSEL}.

\begin{figure}[h]
\includegraphics[scale=0.5]{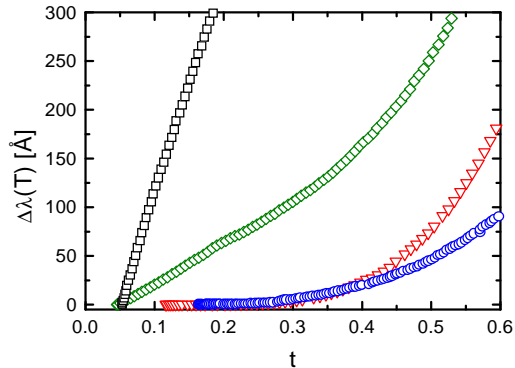}
\caption{Low temperature variation of the in-plane magnetic
penetration depth for different superconductors: a Y124 single
crystal (\small{$\square$}\normalsize), a Y123 single crystal
(\small{$\lozenge$}\normalsize), a NbN epitaxial thin film
(\small{$\bigtriangledown$}\normalsize) and the CaC$_{6}$ c-axis
oriented polycristalline sample under test
(\Large{$\circ$}\normalsize).} \label{Figure4}
\end{figure}

Since the variation of the magnetic penetration depth
$\Delta\lambda(T)$ is proportional to the fraction of normal
electrons in the low-temperature limit, it is useful to compare
for CaC$_{6}$ this quantity versus the reduced temperature
$t=T/T_{c}$ with data taken using the same technique on different
superconductors: an optimally doped
YBa$_{2}$Cu$_{3}$O$_{7-\delta}$ (Y123) single crystal \cite{ERB},
a YBa$_{2}$Cu$_{4}$O$_{8}$ (Y124) single crystal \cite{LAMURAIV}
and an epitaxial NbN thin film \cite{LAMURAII} (fig. 4). In
accordance with the $d$-wave model currently adopted for cuprates,
$\Delta\lambda(T)$ follows a linear dependence for Y123 and Y124
samples. In the case of the $s$-wave BCS conventional
superconductor NbN, the low energy quasiparticle excitation rate
is strongly reduced at low temperatures.  The comparison shows
that CaC$_{6}$ behaves as NbN does, providing a further clear
evidence of the fully gapped nature of superconductivity in this
compound. The ratio $2\Delta(0)/k_{_B}T_{c}$ is in very good
agreement with the BCS theoretical prediction for a conventional
weakly phonon-coupled superconductor: by using the predicted
values for the electron-phonon coupling $\lambda$=0.83 and the
logarithmic averaged phonon frequency $\omega_{ln}=24.7 meV$
\cite{CALANDRA}, the expected ratio is 3.69 \cite{MARSIGLIO}, that
perfectly matches the measured value.

Finally, it is worth to mention that GICs were believed to be an
example of two-band gap superconductivity due to the crossing of
the Fermi surface by both the graphite sheet $\pi$ bands and the
intercalant layer $s$-band \cite{ALJISHI}. However, an attempt to
fit the measurements using a two-band gap model \cite{LAMURAV}
does not provide any additional information simply because a
single-gap analysis is fully capable to describe the data within a
few percent indetermination. This result agrees with the
prediction of Calandra and Mauri \cite{CALANDRA} that the
contribution from the $\pi$ bands is too small to sustain the
superconductivity in CaC$_{6}$. The main contribution seems
therefore to be ascribed to the Ca $s$-band electrons coupled with
C out-of plane and Ca in-plane phononic modes, thus giving origin
to a single gap $s$-wave superconductivity.

In conclusion, we have reported the first measurement of the
magnetic penetration depth in bulk CaC$_{6}$. A standard $s$-wave
BCS model can account for the low temperature experimental data,
allowing a precise estimation of some superconducting parameters
($\Delta(0)=(1.79\pm0.08)$ meV and $\lambda_{ab}(0)=(720\pm80)$
Å). The measured ratio $2\Delta(0)/k_{_B}T_{c}=(3.6\pm0.2)$ is in
excellent agreement with a weakly phonon-coupled model of
superconductivity. These results are in our opinion an important
evidence in favor of a conventional nature of superconductivity in
GICs.

\begin{acknowledgments}
We thank A. Gauzzi for interesting and fruitful discussions. We
are grateful to V. De Luise, A. Maggio and S. Marrazzo for their
valuable technical assistance.
\end{acknowledgments}

$^a$Author to whom correspondence should be addressed. E-mail:
gianrico.lamura@na.infn.it.


\begin{thebibliography}{24}
\expandafter\ifx\csname
natexlab\endcsname\relax\def\natexlab#1{#1}\fi
\expandafter\ifx\csname bibnamefont\endcsname\relax
  \def\bibnamefont#1{#1}\fi
\expandafter\ifx\csname bibfnamefont\endcsname\relax
  \def\bibfnamefont#1{#1}\fi
\expandafter\ifx\csname citenamefont\endcsname\relax
  \def\citenamefont#1{#1}\fi
\expandafter\ifx\csname url\endcsname\relax
  \def\url#1{\texttt{#1}}\fi
\expandafter\ifx\csname
urlprefix\endcsname\relax\def\urlprefix{URL }\fi
\providecommand{\bibinfo}[2]{#2}
\providecommand{\eprint}[2][]{\url{#2}}

\bibitem[{\citenamefont{Weller et~al.}(2005)\citenamefont{Weller, Ellerby,
  Saxena, Smith, and Skipper}}]{WELLER}
\bibinfo{author}{\bibfnamefont{T.~E.} \bibnamefont{Weller}},
  \bibinfo{author}{\bibfnamefont{M.}~\bibnamefont{Ellerby}},
  \bibinfo{author}{\bibfnamefont{A.~S.} \bibnamefont{Saxena}},
  \bibinfo{author}{\bibfnamefont{R.~P.} \bibnamefont{Smith}}, \bibnamefont{and}
  \bibinfo{author}{\bibfnamefont{N.~T.} \bibnamefont{Skipper}},
  \bibinfo{journal}{Nature Phys.} \textbf{\bibinfo{volume}{1}},
  \bibinfo{pages}{39} (\bibinfo{year}{2005}).

\bibitem[{\citenamefont{Emery et~al.}(2005{\natexlab{a}})\citenamefont{Emery,
  Hérold, d'Astuto, V.Garcia, Bellin, Marêché, Lagrange, and Loupias}}]{EMERY}
\bibinfo{author}{\bibfnamefont{N.}~\bibnamefont{Emery}},
  \bibinfo{author}{\bibfnamefont{C.}~\bibnamefont{Hérold}},
  \bibinfo{author}{\bibfnamefont{M.}~\bibnamefont{d'Astuto}},
  \bibinfo{author}{\bibnamefont{V.Garcia}},
  \bibinfo{author}{\bibfnamefont{C.}~\bibnamefont{Bellin}},
  \bibinfo{author}{\bibfnamefont{J.~F.} \bibnamefont{Marêché}},
  \bibinfo{author}{\bibfnamefont{P.}~\bibnamefont{Lagrange}}, \bibnamefont{and}
  \bibinfo{author}{\bibfnamefont{G.}~\bibnamefont{Loupias}},
  \bibinfo{journal}{Phys. Rev. Lett.} \textbf{\bibinfo{volume}{95}},
  \bibinfo{pages}{87003} (\bibinfo{year}{2005}{\natexlab{a}}).

\bibitem[{\citenamefont{Cs\'{a}nyi et~al.}(2005)\citenamefont{Cs\'{a}nyi,
  Littlewood, Nevidomskyy, Pikard, and Simons}}]{CSANYI}
\bibinfo{author}{\bibfnamefont{G.}~\bibnamefont{Cs\'{a}nyi}},
  \bibinfo{author}{\bibfnamefont{P.~B.} \bibnamefont{Littlewood}},
  \bibinfo{author}{\bibfnamefont{A.~H.} \bibnamefont{Nevidomskyy}},
  \bibinfo{author}{\bibfnamefont{C.~P.} \bibnamefont{Pikard}},
  \bibnamefont{and} \bibinfo{author}{\bibfnamefont{B.~D.}
  \bibnamefont{Simons}}, \bibinfo{journal}{Nature Phys.}
  \textbf{\bibinfo{volume}{1}}, \bibinfo{pages}{42} (\bibinfo{year}{2005}).

\bibitem[{\citenamefont{Mazin}(2005)}]{MAZIN}
\bibinfo{author}{\bibfnamefont{I.~I.} \bibnamefont{Mazin}},
  \bibinfo{journal}{Phys. Rev. Lett.} \textbf{\bibinfo{volume}{95}},
  \bibinfo{pages}{227001} (\bibinfo{year}{2005}).

\bibitem[{\citenamefont{Calandra and Mauri}(2005)}]{CALANDRA}
\bibinfo{author}{\bibfnamefont{M.}~\bibnamefont{Calandra}} \bibnamefont{and}
  \bibinfo{author}{\bibfnamefont{F.}~\bibnamefont{Mauri}},
  \bibinfo{journal}{Phys. Rev. Lett.} \textbf{\bibinfo{volume}{95}},
  \bibinfo{pages}{237002} (\bibinfo{year}{2005}).

\bibitem[{\citenamefont{Hannay et~al.}(1965)\citenamefont{Hannay, Geballe,
  Matthias, Andres, Schmidt, and MacNair}}]{HANNAY}
\bibinfo{author}{\bibfnamefont{N.~B.} \bibnamefont{Hannay}},
  \bibinfo{author}{\bibfnamefont{T.~H.} \bibnamefont{Geballe}},
  \bibinfo{author}{\bibfnamefont{B.~T.} \bibnamefont{Matthias}},
  \bibinfo{author}{\bibfnamefont{K.}~\bibnamefont{Andres}},
  \bibinfo{author}{\bibfnamefont{P.}~\bibnamefont{Schmidt}}, \bibnamefont{and}
  \bibinfo{author}{\bibfnamefont{D.}~\bibnamefont{MacNair}},
  \bibinfo{journal}{Phys. Rev. Lett.} \textbf{\bibinfo{volume}{14}},
  \bibinfo{pages}{225} (\bibinfo{year}{1965}).

\bibitem[{\citenamefont{Belash et~al.}(1990)\citenamefont{Belash, Bronnikov,
  Zharikov, and Palnichenko}}]{BELASH}
\bibinfo{author}{\bibfnamefont{I.~T.} \bibnamefont{Belash}},
  \bibinfo{author}{\bibfnamefont{A.~D.} \bibnamefont{Bronnikov}},
  \bibinfo{author}{\bibfnamefont{O.~V.} \bibnamefont{Zharikov}},
  \bibnamefont{and} \bibinfo{author}{\bibfnamefont{A.~V.}
  \bibnamefont{Palnichenko}}, \bibinfo{journal}{Synth. Met.}
  \textbf{\bibinfo{volume}{36}}, \bibinfo{pages}{283} (\bibinfo{year}{1990}).

\bibitem[{\citenamefont{Klapwijk}(2005)}]{KLAPWIJK}
\bibinfo{author}{\bibfnamefont{T.~M.} \bibnamefont{Klapwijk}},
  \bibinfo{journal}{Nature Phys.} \textbf{\bibinfo{volume}{1}},
  \bibinfo{pages}{17} (\bibinfo{year}{2005}).

\bibitem[{\citenamefont{Kasumov et~al.}(2003)\citenamefont{Kasumov, Kociak,
  Ferrier, Deblock, Guéron, Reulet, Khodos, Stéphan, and Bouchiat}}]{KASUMOV}
\bibinfo{author}{\bibfnamefont{A.}~\bibnamefont{Kasumov}},
  \bibinfo{author}{\bibfnamefont{M.}~\bibnamefont{Kociak}},
  \bibinfo{author}{\bibfnamefont{M.}~\bibnamefont{Ferrier}},
  \bibinfo{author}{\bibfnamefont{R.}~\bibnamefont{Deblock}},
  \bibinfo{author}{\bibfnamefont{S.}~\bibnamefont{Guéron}},
  \bibinfo{author}{\bibfnamefont{B.}~\bibnamefont{Reulet}},
  \bibinfo{author}{\bibfnamefont{I.}~\bibnamefont{Khodos}},
  \bibinfo{author}{\bibfnamefont{O.}~\bibnamefont{Stéphan}}, \bibnamefont{and}
  \bibinfo{author}{\bibfnamefont{H.}~\bibnamefont{Bouchiat}},
  \bibinfo{journal}{Phys. Rev. B} \textbf{\bibinfo{volume}{68}},
  \bibinfo{pages}{214521} (\bibinfo{year}{2003}).

\bibitem[{\citenamefont{Emery et~al.}(2005{\natexlab{b}})\citenamefont{Emery,
  Hérold, and Lagrange}}]{EMERYII}
\bibinfo{author}{\bibfnamefont{N.}~\bibnamefont{Emery}},
  \bibinfo{author}{\bibfnamefont{C.}~\bibnamefont{Hérold}}, \bibnamefont{and}
  \bibinfo{author}{\bibfnamefont{P.}~\bibnamefont{Lagrange}},
  \bibinfo{journal}{J. Solid State Chem.} \textbf{\bibinfo{volume}{178}},
  \bibinfo{pages}{2947} (\bibinfo{year}{2005}{\natexlab{b}}).

\bibitem[{\citenamefont{Allender et~al.}(1973)\citenamefont{Allender, Bray, and
  Bardeen}}]{ALLENDER}
\bibinfo{author}{\bibfnamefont{D.}~\bibnamefont{Allender}},
  \bibinfo{author}{\bibfnamefont{J.}~\bibnamefont{Bray}}, \bibnamefont{and}
  \bibinfo{author}{\bibfnamefont{J.}~\bibnamefont{Bardeen}},
  \bibinfo{journal}{Phys. Rev. B} \textbf{\bibinfo{volume}{7}},
  \bibinfo{pages}{1020} (\bibinfo{year}{1973}).

\bibitem[{\citenamefont{Bill et~al.}(2003)\citenamefont{Bill, Morawitz, and
  Kresin}}]{BILL}
\bibinfo{author}{\bibfnamefont{A.}~\bibnamefont{Bill}},
  \bibinfo{author}{\bibfnamefont{H.}~\bibnamefont{Morawitz}}, \bibnamefont{and}
  \bibinfo{author}{\bibfnamefont{V.~Z.} \bibnamefont{Kresin}},
  \bibinfo{journal}{Phys. Rev. B} \textbf{\bibinfo{volume}{68}},
  \bibinfo{pages}{144519} (\bibinfo{year}{2003}).

\bibitem[{\citenamefont{Mazin and Molodtsov}(2005)}]{MAZIN2}
\bibinfo{author}{\bibfnamefont{I.~I.} \bibnamefont{Mazin}} \bibnamefont{and}
  \bibinfo{author}{\bibfnamefont{S.~L.} \bibnamefont{Molodtsov}},
  \bibinfo{journal}{Phys. Rev. B} \textbf{\bibinfo{volume}{72}},
  \bibinfo{pages}{172504} (\bibinfo{year}{2005}).

\bibitem[{\citenamefont{Gauzzi et~al.}(2000)\citenamefont{Gauzzi, Cochec,
  Lamura, Jönsson, Gasparov, Ladan, Plaçais, Probst, Pavuna, and Bok}}]{GAUZZI}
\bibinfo{author}{\bibfnamefont{A.}~\bibnamefont{Gauzzi}},
  \bibinfo{author}{\bibfnamefont{J.~L.} \bibnamefont{Cochec}},
  \bibinfo{author}{\bibfnamefont{G.}~\bibnamefont{Lamura}},
  \bibinfo{author}{\bibfnamefont{B.~J.} \bibnamefont{Jönsson}},
  \bibinfo{author}{\bibfnamefont{V.~A.} \bibnamefont{Gasparov}},
  \bibinfo{author}{\bibfnamefont{F.~R.} \bibnamefont{Ladan}},
  \bibinfo{author}{\bibfnamefont{B.}~\bibnamefont{Plaçais}},
  \bibinfo{author}{\bibfnamefont{P.~A.} \bibnamefont{Probst}},
  \bibinfo{author}{\bibfnamefont{D.}~\bibnamefont{Pavuna}}, \bibnamefont{and}
  \bibinfo{author}{\bibfnamefont{J.}~\bibnamefont{Bok}}, \bibinfo{journal}{Rev.
  Sci. Instr.} \textbf{\bibinfo{volume}{71}}, \bibinfo{pages}{2147}
  (\bibinfo{year}{2000}).

\bibitem[{\citenamefont{Lamura et~al.}(2003)\citenamefont{Lamura, Cochec,
  Gauzzi, Licci, Castro, Bianconi, and Bok}}]{LAMURAIII}
\bibinfo{author}{\bibfnamefont{G.}~\bibnamefont{Lamura}},
  \bibinfo{author}{\bibfnamefont{J.~L.} \bibnamefont{Cochec}},
  \bibinfo{author}{\bibfnamefont{A.}~\bibnamefont{Gauzzi}},
  \bibinfo{author}{\bibfnamefont{F.}~\bibnamefont{Licci}},
  \bibinfo{author}{\bibfnamefont{D.~D.} \bibnamefont{Castro}},
  \bibinfo{author}{\bibfnamefont{A.}~\bibnamefont{Bianconi}}, \bibnamefont{and}
  \bibinfo{author}{\bibfnamefont{J.}~\bibnamefont{Bok}},
  \bibinfo{journal}{Phys. Rev. B} \textbf{\bibinfo{volume}{67}},
  \bibinfo{pages}{144518} (\bibinfo{year}{2003}).

\bibitem[{\citenamefont{Turneaure et~al.}(1991)\citenamefont{Turneaure,
  Halbritter, and Schwettman}}]{TURNEAURE}
\bibinfo{author}{\bibfnamefont{J.~P.} \bibnamefont{Turneaure}},
  \bibinfo{author}{\bibfnamefont{J.}~\bibnamefont{Halbritter}},
  \bibnamefont{and} \bibinfo{author}{\bibfnamefont{H.~A.}
  \bibnamefont{Schwettman}}, \bibinfo{journal}{J. Supercond.}
  \textbf{\bibinfo{volume}{4}}, \bibinfo{pages}{341} (\bibinfo{year}{1991}).

\bibitem[{\citenamefont{Tinkham}(1996)}]{TINK}
\bibinfo{author}{\bibfnamefont{M.}~\bibnamefont{Tinkham}},
  \emph{\bibinfo{title}{Introduction to superconductivity}}
  (\bibinfo{publisher}{McGraw-Hill}, \bibinfo{address}{New York},
  \bibinfo{year}{1996}).

\bibitem[{\citenamefont{Dresselhaus and Dresselhaus}(2000)}]{DRESSEL}
\bibinfo{author}{\bibfnamefont{M.~S.} \bibnamefont{Dresselhaus}}
  \bibnamefont{and}
  \bibinfo{author}{\bibfnamefont{G.}~\bibnamefont{Dresselhaus}},
  \bibinfo{journal}{Adv. Phys.} \textbf{\bibinfo{volume}{51}},
  \bibinfo{pages}{1} (\bibinfo{year}{2000}).

\bibitem[{\citenamefont{Erb et~al.}(1995)\citenamefont{Erb, Walker, and
  Flükiger}}]{ERB}
\bibinfo{author}{\bibfnamefont{A.}~\bibnamefont{Erb}},
  \bibinfo{author}{\bibfnamefont{E.}~\bibnamefont{Walker}}, \bibnamefont{and}
  \bibinfo{author}{\bibfnamefont{R.}~\bibnamefont{Flükiger}},
  \bibinfo{journal}{Physica C} \textbf{\bibinfo{volume}{245}},
  \bibinfo{pages}{245} (\bibinfo{year}{1995}).

\bibitem[{\citenamefont{Lamura et~al.}(2006)\citenamefont{Lamura, Gauzzi,
  Kazakov, Karpinski, and Andreone}}]{LAMURAIV}
\bibinfo{author}{\bibfnamefont{G.}~\bibnamefont{Lamura}},
  \bibinfo{author}{\bibfnamefont{A.}~\bibnamefont{Gauzzi}},
  \bibinfo{author}{\bibfnamefont{S.}~\bibnamefont{Kazakov}},
  \bibinfo{author}{\bibfnamefont{J.}~\bibnamefont{Karpinski}},
  \bibnamefont{and} \bibinfo{author}{\bibfnamefont{A.}~\bibnamefont{Andreone}},
  \bibinfo{journal}{J. Phys. Chem. Solids}  (\bibinfo{year}{2006}),
  \bibinfo{note}{in press}.

\bibitem[{\citenamefont{Lamura et~al.}(2002)\citenamefont{Lamura, Villégier,
  Gauzzi, Cochec, Laval, Plaçais, Hadacek, and Bok}}]{LAMURAII}
\bibinfo{author}{\bibfnamefont{G.}~\bibnamefont{Lamura}},
  \bibinfo{author}{\bibfnamefont{J.~C.} \bibnamefont{Villégier}},
  \bibinfo{author}{\bibfnamefont{A.}~\bibnamefont{Gauzzi}},
  \bibinfo{author}{\bibfnamefont{J.~L.} \bibnamefont{Cochec}},
  \bibinfo{author}{\bibfnamefont{J.-Y.} \bibnamefont{Laval}},
  \bibinfo{author}{\bibfnamefont{B.}~\bibnamefont{Plaçais}},
  \bibinfo{author}{\bibfnamefont{N.}~\bibnamefont{Hadacek}}, \bibnamefont{and}
  \bibinfo{author}{\bibfnamefont{J.}~\bibnamefont{Bok}},
  \bibinfo{journal}{Phys. Rev. B} \textbf{\bibinfo{volume}{65}},
  \bibinfo{pages}{104507} (\bibinfo{year}{2002}).

\bibitem[{\citenamefont{Marsiglio et~al.}(1990)\citenamefont{Marsiglio,
  Carbotte, and Blezius}}]{MARSIGLIO}
\bibinfo{author}{\bibfnamefont{F.}~\bibnamefont{Marsiglio}},
  \bibinfo{author}{\bibfnamefont{J.~P.} \bibnamefont{Carbotte}},
  \bibnamefont{and} \bibinfo{author}{\bibfnamefont{J.}~\bibnamefont{Blezius}},
  \bibinfo{journal}{Phys. Rev. B} \textbf{\bibinfo{volume}{41}},
  \bibinfo{pages}{6457} (\bibinfo{year}{1990}).

\bibitem[{\citenamefont{Al-Jishi}(1983)}]{ALJISHI}
\bibinfo{author}{\bibfnamefont{R.}~\bibnamefont{Al-Jishi}},
  \bibinfo{journal}{Phys. Rev. B} \textbf{\bibinfo{volume}{28}},
  \bibinfo{pages}{28} (\bibinfo{year}{1983}).

\bibitem[{\citenamefont{Lamura et~al.}(2001)\citenamefont{Lamura, Gennaro,
  Salluzzo, Andreone, Cochec, Gauzzi, Cantoni, Paranthaman, Christen, Christen
  et~al.}}]{LAMURAV}
\bibinfo{author}{\bibfnamefont{G.}~\bibnamefont{Lamura}},
  \bibinfo{author}{\bibfnamefont{E.~D.} \bibnamefont{Gennaro}},
  \bibinfo{author}{\bibfnamefont{M.}~\bibnamefont{Salluzzo}},
  \bibinfo{author}{\bibfnamefont{A.}~\bibnamefont{Andreone}},
  \bibinfo{author}{\bibfnamefont{J.~L.} \bibnamefont{Cochec}},
  \bibinfo{author}{\bibfnamefont{A.}~\bibnamefont{Gauzzi}},
  \bibinfo{author}{\bibfnamefont{C.}~\bibnamefont{Cantoni}},
  \bibinfo{author}{\bibfnamefont{M.}~\bibnamefont{Paranthaman}},
  \bibinfo{author}{\bibfnamefont{D.~K.} \bibnamefont{Christen}},
  \bibinfo{author}{\bibfnamefont{H.~M.} \bibnamefont{Christen}},
  \bibnamefont{et~al.}, \bibinfo{journal}{Phys. Rev. B}
  \textbf{\bibinfo{volume}{65}}, \bibinfo{pages}{20506} (\bibinfo{year}{2001}).

\end{thebibliography}
\end{document}